\documentclass[pre,twocolumn,groupedaddress,showpacs,nofootinbib]{revtex4}
\usepackage{graphicx,amsmath,amssymb}
\usepackage{bm}
\usepackage{color}
\newcommand{\potts}{\text{Potts}}
\newcommand{\axel}{\text{Axelrod}}
\newcommand{\nen}{\langle ij\rangle}
\newcommand{\Eq}[1]{Eq.\@ (\ref{#1})}
\newcommand{\Fig}[1]{Fig.\@ \ref{#1}}
\newcommand{\Figs}[1]{Figs.\@ \ref{#1}}
\newcommand{\Sec}[1]{Sec.\@ \ref{#1}}
\newcommand{\Ref}[1]{Ref.\@ \cite{#1}}
\newcommand{\App}[1]{App.\@ \ref{#1}}

\begin{document}
\title{Nature of phase transitions in Axelrod-like coupled Potts models in two dimensions} 
\author{Yerali Gandica}
\email{ygandica@gmail.com}
\affiliation{Department of Mathematics and Namur Center for Complex Systems - naXys, University of Namur, 
rempart de la Vierge 8, B 5000 Namur, Belgium}
\author{Silvia Chiacchiera}
\email{silvia@teor.fis.uc.pt}
\affiliation{CFisUC, Department of Physics, University of Coimbra, P3004-516 Coimbra, Portugal}
\date{\today}
\begin{abstract}
We study $F$ coupled $q$-state Potts models in a two-dimensional square lattice. 
The interaction between the different layers is attractive, to favour a
simultaneous alignment in all of them, and its strength is fixed.
The nature of the phase transition for zero field is numerically determined for $F=2,3$. 
Using the Lee-Kosterlitz method, we find that it is continuous for $F=2$ and  $q=2$, whereas it is abrupt
 for higher values of $q$ and/or $F$.
When a continuous or a weakly first-order phase transition takes place, we also analyze the properties 
of the geometrical clusters. This allows us to determine the fractal dimension $D$ of the incipient 
infinite cluster and to examine the finite-size scaling of the cluster number density via data collapse.
A mean-field approximation of the model, from which some general trends can be determined, is presented too.
Finally, since this lattice model has been recently considered as a thermodynamic counterpart of the Axelrod
 model of social dynamics, we discuss our results in connection with this one.
\end{abstract}
\maketitle
\section{Introduction}
Given a model of statistical mechanics whose thermodynamic behaviour and phase diagram 
are known, one can ask oneself what is the macroscopic behaviour that 
sets in when two or more such models are coupled together. 
For example, one motivation for this question could be to understand
the effect of randomness, or, simply,
the composed model could be interesting by itself. 
In general, when two or more models are coupled, the symmetry of a system and, consequently, its
critical behaviour \cite{kardar,goldenfeld,stanley,pathria}, are altered. Even the phase transition
 nature may change. However, {\it a priori} it is not possible to say how dramatic these changes will be,
 or how sensitive they are to the choice of the inter-model coupling. 

The $q$-state Potts model \cite{Potts52,Wu82} is a very popular lattice model which has been extensively 
studied within many different theoretical approaches. On the experimental side, systems from condensed 
matter obeying the same symmetry have been identified.
For a two-dimensional square lattice geometry, it is known that a single $q$-state Potts model, 
in absence of external fields,  undergoes a temperature-driven phase transition whose nature changes 
with $q$: it is continuous for $q\le4$ and abrupt for $q>4$. 
Concerning the critical behaviour, explicit forms have been conjectured for the thermal \cite{denNijs79}
and magnetic \cite{Nienhuis82} exponents for $q\le4$, from which all the thermodynamic critical exponents can be derived.
These conjectured values seem to be generally confirmed by numerical and perturbative approaches \cite{Wu82}.
Besides the thermal behaviour, one can analyze the geometric one. 
The basic geometric object to consider in a lattice model is a cluster, a connected set of nearest 
neighbours in the same state. Varying the temperature, one typically observes that a geometrical phase transition, 
analogous to percolation, takes place \cite{stauffer,fortunato02}. 
There are critical exponents associated with this transition too, 
and, in principle, they are independent of the thermal ones. In this line, for the Potts model, 
closed forms for the various fractal dimensions (bulk, surface, etc) of the infinite cluster at criticality
have been obtained in \Ref{Vanderzande92}.

In this work we study how the phase transition type and the critical behaviour of the $q$-state 
Potts model in a two-dimensional square lattice change when $F$ similar models (layers) are coupled. 
In particular, we adopt a four-point coupling of fixed strength between the layers.
Our aim is to see how the known behaviour and the critical properties of the single-layer model 
are altered by the coupling.

Coupled Potts models have been discussed in the literature, mainly to get insight on a single model 
with random couplings \cite{Dotsenko99}. 
Matsuda \cite{Matsuda81} considered two Potts models coupled with a four-body interaction; in particular,
he located the transition temperature in the two-dimensional case using self-duality arguments.
Later, he generalized the study to many coupled layers for a Hamiltonian containing up to $2F$-point terms 
\cite{Matsuda83}.
More recently, Dotsenko {\it et al.} \cite{Dotsenko99} considered the fixed-point structure and stability for
a similar Hamiltonian to identify the corresponding conformal field theory.
For the case $q=2$, our model is equivalent to the $F$-colour Ashkin-Teller (AT) model \cite{Ashkin43,Baxter,Grest81},
 which in two dimensions has been studied for a weak four-body coupling and generic $F$ in \Ref{Grest81} and in the limit 
$F\rightarrow \infty$ in \Ref{Fradkin1984}. Throughout the article, we will make comparisons with previous results 
whenever possible.

Beyond its relevance for the field of critical phenomena, the study developed in this article is also motivated
 by a possible connection to a model of social dynamics.
At the end of the 1990s, Axelrod introduced an agent-based vectorial model for the spread of 
social influence \cite{Axelrod97}. In this non-equilibrium model the agents live on a lattice and 
are characterized by a dynamical culture, a set of $F$ features each assuming values out 
of a finite set of $q$ traits. The interactions are local and act to increase the similarity between agents.
Systematic quantitative studies of the Axelrod model have been performed, exploring different aspects: 
non-equilibrium phase transitions \cite{Castellano00}, the effects of noise \cite{Klemm03,Klemm05}, 
dimensionality \cite{Klemm03a} and size finiteness \cite{Toral07}. 
Keeping in mind the delicacy of the connection between social and physical phenomena, 
thermodynamic (equilibrium) counterparts of the Axelrod model have been proposed 
\cite{Gandica13,Genzor14}. In particular, the Hamiltonian we analyze in this work coincides with that of 
\Ref{Gandica13}, in which the authors focused on the one-dimensional case. 

This article is organized as follows. In \Sec{sec:model} we present the model and relate it to similar
 ones from statistical mechanics. Its possible connection to the Axelrod non-equilibrium model of 
social dynamics is briefly recalled too. 
In \Sec{Sec:MF} a mean-field approximation is carried out, and the findings 
for phase transitions at this level are summarized. In the subsequent sections, we present the results
 of a numerical Monte Carlo study on a two-dimensional square lattice, for different choices of $F$ 
and $q$. In particular, in \Sec{sec:phasetr} we analyze the phase transition type. 
For the cases $F=2, q=2$ and $F=3, q=2$, where continuous and weakly first-order phase 
transitions take place, respectively,
we analyze in \Sec{sec:perco} the geometrical critical (or pseudocritical) behaviour and determine the fractal dimension 
of the incipient infinite cluster. Finally, \Sec{sec:conclu} is devoted to our conclusions. 
\section{The model: $F$ coupled $q$-state Potts models, or, Thermodynamic Axelrod} \label{sec:model}
Our model, as in \Ref{Gandica13}, is constituted by $F$ coupled layers of $q$-state Potts models 
\cite{Potts52,Wu82}. 
The variable at the $i$th site of the lattice is an $F$-vector 
$\vec{\sigma}_i=(\sigma_{i1}, \sigma_{i2},\dots,\sigma_{iF})$, and each of its components has $q$ 
possible values (we choose $0, 1, \dots,q-1$). The Hamiltonian for $N$ sites 
in presence of an external field reads
\begin{eqnarray}
\mathcal{H} &=&
-\sum_{k=1}^F \sum_{\nen} J_{ij}\,\delta(\sigma_{ik},\sigma_{jk}) 
- H \sum_{k=1}^F \sum_{i=1}^N \delta(\sigma_{ik},H_{k})~,\nonumber\\
&&\quad\text{where}\quad J_{ij}=\sum_{l=1}^F J~ \delta(\sigma_{il},\sigma_{jl}) \,,
\label{hamil}
\end{eqnarray}
$J>0$ and $H\ge0$ are constants, and $\delta(\cdot\,,\cdot)$ is Kronecker's delta. 
The symbol $\langle\cdot \rangle$ indicates nearest neighbour sites, and the sum is over distinct pairs.
The quantities ${H_k}$ are the favoured directions selected by the external field.
Rephrased in the language of the social Axelrod model, each agent, which may be seen as representing a person or a small village, 
is characterized by a dynamical \emph{culture}, a set of ($F$) \emph{features} each assuming values
 out of a finite set of ($q$) \emph{traits}. The interaction in \Eq{hamil} is built to reproduce the Axelrod rules: 
the energy of a link decreases with the interaction strength (as it is common in Hamiltonian systems)
and the interaction strength increases when the agents are aligned in the same direction. 
Besides similarities, there is an important difference between the original Axelrod model and this thermodynamic 
version: in the Axelrod model, the interactions between two agents stop when they are identical or completely different, 
and the dynamics leads to absorbing configurations. That is not the case in the thermodynamic version, where 
thermal fluctuations are present. In this sense, our model, which never gets frozen, should rather be compared
 with the Axelrod model in the presence of noise \cite{Klemm03,Klemm05}.

For the case of one layer ($F=1$), \Eq{hamil} reduces to the standard $q$-state Potts model. 
When more layers are present, our model is constituted by $F$ layers of $q$-state Potts models coupled 
{\it via} their energies (i.e., with a four-body term) \cite{Matsuda81,Dotsenko99}. 
When comparing with similar models, notice that ours contains only one coupling constant ($J$) 
and the ratio between the strengths of the two-body and four-body terms is fixed.
This is clearly seen by rewriting \Eq{hamil} as follows
\begin{widetext}
\begin{equation}\mathcal{H} =
-J \sum_{k=1}^F  \sum_{\nen} \delta(\sigma_{ik},\sigma_{jk}) 
-J \sum_{\substack{k,l=1\\ k\ne l}}^F  \sum_{\nen} \delta(\sigma_{ik},\sigma_{jk}) 
\delta(\sigma_{il},\sigma_{jl}) 
- H \sum_{k=1}^F \sum_{i=1}^N \delta(\sigma_{ik},H_{k}) 
. \label{hamil2}
\end{equation}
\end{widetext}
For the case $q=2$, our model is equivalent to the $F$-colour Ashkin-Teller model 
\cite{Ashkin43,Baxter,Grest81} with (in Grest's notation) a two-body coupling $K_2=\beta J F/2$, 
and the constraint $K_4/K_2=1/F$ between the two- and four- body couplings, as 
can easily be checked by rewriting \Eq{hamil} in terms of a spin-$\frac{1}{2}$ variable, assuming values $\pm 1$.
\section{Mean-Field approximation}\label{Sec:MF}
In a Mean-Field (MF) approximation, an interacting system is replaced by independent constituents embedded 
in a modified environment. In our case, each agent, instead of interacting with its nearest
neighbours, feels a uniform field, originated by the average orientation of the others.
Many different recipes can be used to build such an approximation. In this section we present an 
approach constructed at the level of the probabilities for the various states 
\cite{LoicTurban,Nishimori2011}.
\subsection{Probabilities and variational method}
In statistical mechanics, equilibrium averages of a generic quantity $A$ are obtained by taking into
 account all the possible microstates $\mu_s$, each with an appropriate weight $p(\mu_s)$:
\begin{equation}
\langle A\rangle=\sum_{\mu_s}p(\mu_s)A(\mu_s)~,
\end{equation}
where the normalization condition $\sum_{\mu_s}p(\mu_s)=1$ is satisfied.
The specific form of the probabilities depends on the macroscopic constraints imposed on the system. 
We work in the canonical ensemble, where the weights are computed from the microscopic 
Hamiltonian ${\mathcal H}$ as $p(\mu_s)=e^{-\beta {\mathcal H}(\mu_s)}/Z$, 
and the normalization factor is the partition function $Z=\sum_{\mu_s}e^{-\beta {\mathcal H}(\mu_s)}$.
As usual, $\beta=1/k_BT$ is the inverse temperature, and $k_B$ the Boltzmann's constant.
All the thermodynamics can be extracted from the partition function $Z$, 
through its relation to the Gibbs free energy $G$, namely $G=-k_BT \ln{Z}$.

Consider now a generic probability $p_{0}$: the following inequality can be easily shown 
\begin{equation}\label{eq:G0}
G\le G_0=\langle \mathcal{H}\rangle_0+k_BT\langle \ln p_0\rangle_0= E_0 -TS_0~,
\end{equation}
where $\langle \dots \rangle_0$ denotes the averages computed using the probability $p_0$.
The average energy $E_0=\langle {\mathcal H}\rangle_0$ and the statistical entropy 
$S_0=-k_B\langle \ln p_0\rangle_0$, both obtained using $p_0$,
 have been introduced too. 
\Eq{eq:G0} shows that any free energy $G_0$ computed in this way will be larger than the exact Gibbs
 free energy. In the spirit of variational methods, a form is proposed for $p_0$, which will depend 
upon a number of parameters; the best approximation to the exact free energy is then 
obtained minimizing $G_0$ with respect to these parameters.

In a MF approximation, the system is treated at a single-particle level. Therefore, the probability
 function factorizes into $N$ terms, one for each site,
\begin{equation}
p(\vec{\sigma_1},\dots,\vec{\sigma_N})\simeq  \prod_{i=1}^N p_i(\vec{\sigma_i})~
\quad\text{(MF approximation)}.
\end{equation}
We underline that in this approach the full Hamiltonian is used; there are other MF 
approaches in which, instead, the simplifications are done at the level of the Hamiltonian 
\cite{Nishimori2011}. In general, the physical results are the same.
\subsection{Mean-field approximation for the Thermodynamic Axelrod model}
We present in this section the MF approximation of our model \Eq{hamil}.
The calculations are done for a lattice of generic dimensionality and geometry,
and coordination number $z$, since this is the only lattice property affecting the MF results.
In \App{app:MFpotts} we give a more detailed derivation for the case of one layer ($F=1$). 

For symmetry reasons all the features behave in the same way, then the average
occupation of the trait $\alpha$ for the feature (=layer) $k$ is
\begin{equation}\label{eq:munu}
\langle \delta(\sigma_{ik},\alpha)\rangle=\left\{
\begin{array}{l}
\mu, \quad\text{if}\quad\alpha=0, \forall k\\
\nu, \quad\text{if}\quad\alpha\neq0, \forall k~.
\end{array}
\right.
\end{equation}
In the latter we have used translational invariance, and, without loss of generality, imagined
that the external field favours the trait $0$ in any feature
 ($H_k=0, \forall k$). Of course, $\mu$ and $\nu$ are related by the constraint $\mu+(q-1)\nu=1$. 
It is convenient to define as an order parameter 
\begin{equation}\label{eq:OPmf}
m=\mu-\nu~\quad\quad\text{(MF order parameter)}:
\end{equation} 
as $T\rightarrow 0$, we expect $\mu=1$ and $\nu=0$, then $m=1$;
as $T\rightarrow \infty$, we expect $\mu=\nu=1/q$, then $m=0$.

In MF approximation, we assume that the full probability factorizes into $N\times F$ terms
\begin{equation}
p_{MF}(\vec{\sigma_1},\dots,\vec{\sigma_N})=\prod_{i=1}^N p_{i}(\vec{\sigma_{i}})
=\prod_{i=1}^N \prod_{k=1}^F p_{ik}(\sigma_{ik})~.
\end{equation}
Notice that, with this choice, the different components at a site 
are not correlated. 
Throughout this study, both at the MF level and in the numerical simulations, 
we only distinguish between a (completely) ordered phase, in which all the components are
ordered, and a disordered phase, in which they are all disordered. 
Phases in which the orientation is random within any layer, but 
at a single site the variables are correlated,
are known to appear in the AT model 
when the four-body coupling is considerably larger than the one considered here 
\cite{Baxter,Grest81}.
For any layer we assume the following structure of the probability 
\begin{equation}
p_{ik}(\sigma_{ik})=\frac{1+m[q\, \delta(\sigma_{ik},0)-1]}{q}~,\quad \sigma_{ik}=0,1,\dots,q-1~,
\end{equation}
which is derived in detail for the $q$-state Potts model in \App{app:MFpotts} 
(see also \Ref{LoicTurban}).

Defining  $\Delta_{ij}=\left[\sum_{k=1}^F \delta(\sigma_{ik},\sigma_{jk})\right]^2$,
one can rewrite the interaction term in \Eq{hamil} as $-J\sum_{\nen}\Delta_{ij}$.
In view of the MF, is convenient to elaborate $\Delta_{ij}$
so that it becomes a product of terms, each depending on a single component at a site: using 
\Eq{eq:exdeltas} it becomes 
\begin{eqnarray}
\Delta_{ij}&=&\sum_{\substack{k,l=1\\ k\ne l}}^F\sum_{\alpha,\beta=0}^{q-1}  
\delta(\sigma_{ik},\alpha)\delta(\sigma_{jk},\alpha)
\delta(\sigma_{il},\beta)\delta(\sigma_{jl},\beta)\nonumber\\
&&+\sum_{k=1}^F\sum_{\alpha=0}^{q-1}  \delta(\sigma_{ik},\alpha)\delta(\sigma_{jk},\alpha)~.
\end{eqnarray}
Taking the averages we get
\begin{eqnarray}
\langle\Delta_{ij}\rangle_{MF}=
\sum_{\substack{k,l=1\\ k\ne l}}^F[\mu^2+(q-1)\nu^2]^2+\sum_{k=1}^F[\mu^2+(q-1)\nu^2]\nonumber\\
=F(F-1)\left[\frac{1+m^2(q-1)}{q}\right]^2+ F\left[\frac{1+m^2(q-1)}{q}\right]~.\nonumber\\
\end{eqnarray}
The external field term at one site is simply
$\langle\sum_{k=1}^F\delta(\sigma_{ik},H_{k})\rangle_{MF}=F\mu=F [1+m(q-1)]/q$~.
For the statistical entropy, we must recall that now at any site we have $F$ variables, each of which can be in state $0$ 
or not. For symmetry reasons, we are only interested to know how many components 
are not in state $0$: all the others will be in state $0$. 
The number of ways of having $k$ of the $F$ components 
different from state $0$ is $\binom{F}{k}(q-1)^k$,
the weight of each of them is computed as the product 
$\nu^k\mu^{F-k}$. As a check, notice that 
$\sum_{k=0}^F\binom{F}{k}(q-1)^k=q^F$, the total number 
of states for a site, as it should.
One can directly write
\begin{eqnarray}
&&\langle \ln{p_{MF}}\rangle_{MF} =\nonumber\\
&=& N\sum_{k=0}^F
\binom{F}{k}
(q-1)^k ~\nu^k ~\mu^{F-k}\ln{\left(\nu^k ~\mu^{F-k}\right)}\nonumber\\
&=&N\frac{1}{q^F}\sum_{k=0}^F
\binom{F}{k}(q-1)^k(1-m)^k\left[1+m(q-1)\right]^{F-k}\nonumber\\
&&\times\ln\left\{\frac{(1-m)^k\left[1+m(q-1)\right]^{F-k}}{q^F}\right\}~.\quad\quad\quad
\end{eqnarray}

Combining all the contributions together we obtain the free energy per particle of Axelrod model in MF approximation
\begin{widetext}
\begin{eqnarray}\label{eq:MFaxel}
g_{MF}^\axel(m,T) &\equiv&\frac{G_{MF}}{N}=
\frac{\langle \mathcal{H}\rangle_{MF} +k_BT\langle \ln{p_{MF}}\rangle_{MF}}{N} \nonumber\\
&=&-\frac{Jz}{2}\left\{F(F-1)\left[\frac{1+m^2(q-1)}{q}\right]^2+ F\left[\frac{1+m^2(q-1)}{q}\right]\right\}
-HF\frac{1+m(q-1)}{q}\nonumber\\
&&+\frac{k_BT}{q^F}\sum_{k=0}^F
\binom{F}{k}(q-1)^k(1-m)^k\left[1+m(q-1)\right]^{F-k}
\ln\left\{\frac{(1-m)^k\left[1+m(q-1)\right]^{F-k}}{q^F}\right\}~.
\end{eqnarray}
\end{widetext}
Notice that, as it should, \Eq{eq:MFaxel} reduces to \Eq{eq:MFpotts} in the case $F=1$.
\subsection{Phase transitions at the mean-field level}
Following the evolution of the global minimum of $g_{MF}^\axel(m,T)$ with the temperature, we can
 locate possible phase transitions and determine their type. 
In all the cases we analyzed, there is a transition from a low-temperature ordered phase ($m\neq 0$)
 to a high-temperature disordered one ($m=0$). 
For $F=1$ we reproduce the known result that the phase transition, at the mean-field level, is continuous
 for $q=2$ and abrupt for $q>2$ (see also \App{app:MFpotts}). For $F>1$, we find that the transition, at the mean-field level,
 is abrupt for any value of $q\ge 2$. 
Concerning the transition temperatures, we find the following systematics: for a given $F$, the
 transition temperature decreases as $q$ increases; for a given $q$, the transition temperature 
increases as $F$ increases. 
This can be understood as follows: increasing $q$ favours the entropic term, while increasing $F$ 
strengthens the interaction term (see \Eq{hamil}). 

\begin{figure}[t]
\begin{center}
\includegraphics[height=5cm,angle=0]{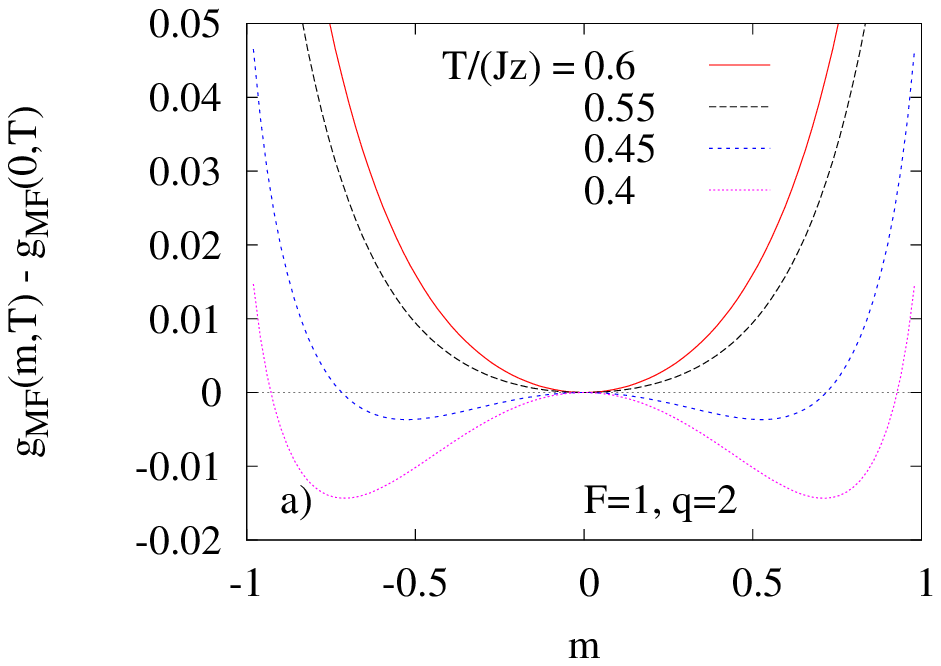}
\includegraphics[height=5cm,angle=0]{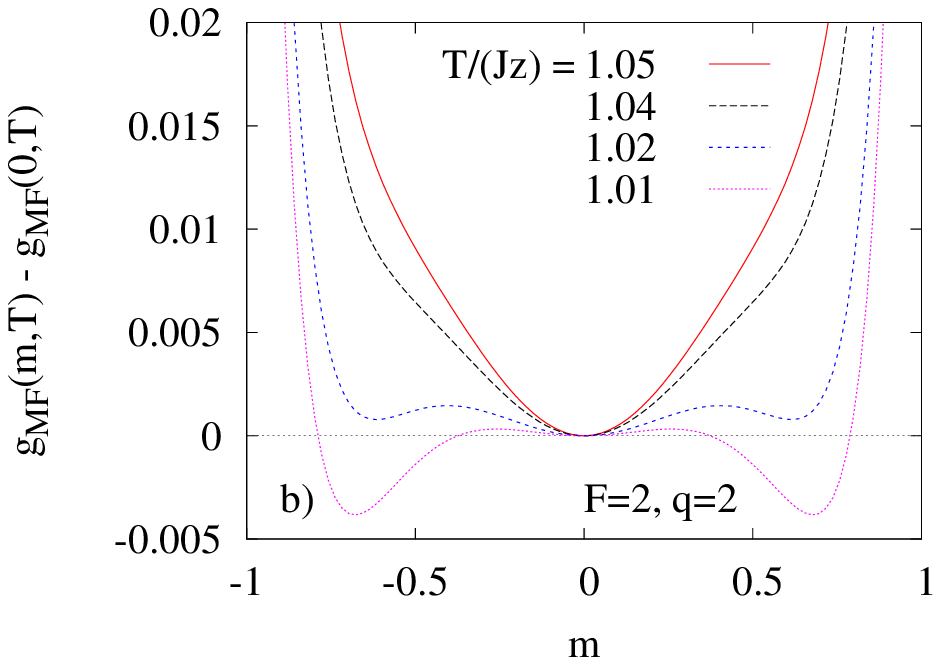}
\end{center}
\caption{Free energy per particle of the thermodynamic Axelrod model 
in MF approximation, \Eq{eq:MFaxel},
 for $H=0$, as a function of the order parameter $m$. 
(a) $F=1,q=2$; (b) $F=2,q=2$. In both cases, the function is shown for four 
values of the temperature in a range which encompasses the phase transition.
}
\label{fig:Fmf}
\end{figure}

\begin{figure}[t]
\begin{center}
\includegraphics[height=5cm,angle=0]{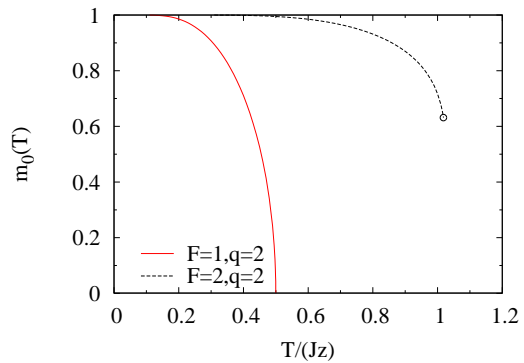}
\end{center}
\caption{Physical value of the order parameter $m$ in the MF approximation [obtained as the global minimum of 
the free energy \Eq{eq:MFaxel}], for $H=0$, as a function of the temperature, for the same two choices 
of $F$ and $q$ used in \Fig{fig:Fmf}.}
\label{fig:OPmf}
\end{figure}

As representative examples, in \Fig{fig:Fmf}
we show the mean-field free energy for the cases $F=1,q=2$ [\Fig{fig:Fmf}(a)] and 
$F=2,q=2$ [\Fig{fig:Fmf}(b)]. In the first one, a continuous phase transition 
takes place, whereas in the second one it is abrupt. 
This can be better seen in \Fig{fig:OPmf}
where the global minimum $m_0$ is shown as a function of the temperature and for the same 
two choices of $F$ and $q$.

Since at the MF level the fluctuations, whose strength is responsible for the singularities 
characterizing second-order phase transitions, are suppressed, it is natural to expect 
that some of the transitions appear abrupt at the MF level as an effect of the approximation.
\section{Phase transition type}\label{sec:phasetr}
In this section we present our results for the phase transition type, obtained analyzing 
Monte Carlo simulations {\it via} the Lee-Kosterlitz method \cite{Lee90,Gould96}. 
Before discussing the results obtained for different choices of $F$ and $q$, we briefly
 recall the method used.

The Lee-Kosterlitz method allows to predict, from relatively small lattices, 
the nature of the phase transition taking place in the system in the thermodynamic limit. 
The central idea of this numerical approach is to monitor the evolution of the quantity 
$\Delta f(L)$, the height of the free energy barrier between two degenerate minima, 
as the system linear size, $L$, increases. If $\Delta f(L)$ increases with $L$, the transition is abrupt;
 if $\Delta f(L)$ is constant the transition is continuous; if $\Delta f(L)$ decreases there
 is no transition in the thermodynamic limit.
The energy barrier is computed on a function $f(X,L)$ which is a restricted free energy, 
including only configurations which 
share the same value of the observable $X$, whose choice depends on the nature of the transition. 
For field-driven transitions, it is the order parameter ($X=M$), while for temperature-driven ones it is the energy 
($X=E$). As usual, to reduce the computational costs, we use this method in combination with the histogram method
by Ferrenberg and Swendsen \cite{Ferrenberg88}. 
The latter allows one to use a single Monte Carlo simulation at a given temperature to obtain
estimates on the system thermodynamic quantities at temperatures nearby. 
Before going on, a warning is in order on the Lee-Kosterlitz method. 
While an increasing $\Delta f(L)$ is a (quite) safe proof of an abrupt phase transition, 
other cases are more tricky. In fact, the method relies on 
the behaviour of the energy barrier between two minima, but it can 
happen that this peak structure is not visible: in such a case, the transition may be continuous, or, 
it may be that the considered values of $L$ are too small to detect the barrier numerically, because 
the transition is weakly abrupt \cite{Kosterlitz93}.

Next we briefly summarize the numerical recipe used for a generic magnetic model.
One starts with a long simulation at a given temperature ($k_BT_0=1/\beta_0$) 
close to the expected transition temperature, and 
builds the double histogram $H_0(E,M;L)$, counting the realizations with energy $E$ 
and (scalar) magnetization $M$. With this, the two restricted free 
energies are obtained as
\begin{equation}
f(E,\beta;L)=-\ln[\sum_M H_0(E,M;L) e^{-(\beta-\beta_0)E}]~,
\end{equation}
and
\begin{equation}
f(M,\beta;L)=-\ln[\sum_E H_0(E,M;L) e^{-(\beta-\beta_0)E}]~.
\end{equation}
In the latter, we have included $\beta$ explicitly and written $f$ as a function
 of three variables. For the sake of clarity in the text and in the plots, when there is no ambiguity,
 we use the compact notation $f(E)$ and $f(M)$ to indicate these quantities.
 
For field-driven transitions, one fixes the temperature and lattice size, and measures the barrier 
height between degenerate minima in the function $f(M,\beta;L)$  
(corresponding to ordering along different directions). The procedure is repeated 
for lattices of different sizes and a set of temperatures.
Instead, for temperature-driven transitions, for any size $L$, one looks for the temperature 
at which $f(E,\beta;L)$ has two degenerate minima (corresponding to ordered and
 disordered phases). The barrier height between these minima is computed for different 
sizes.

We now come to our results. We have performed a numerical Monte Carlo study of the Hamiltonian 
(\ref{hamil}) in a two-dimensional square lattice with periodic boundary conditions,
 using the Metropolis algorithm. The lattices are $L\times L$, with $N=L^2$ sites. 
A common definition of the order parameter to be used in simulations 
for the $q$-state Potts model is \cite{Challa86}
\begin{equation}
m_{MC, F=1}=\frac{q x_{maj}-1}{q-1}~,
\end{equation}
where $x_{maj}$ is the average fractional occupation of the majority 
state\footnote{Notice that this definition coincides with that given in MF, 
\Eq{eq:munu} and \Eq{eq:OPmf}, if all the minority states are all equally occupied. 
In a numerical simulation, this is true only on average.}. 
For a generic $F$, we use a similar definition
\begin{equation}
m_{MC}=\frac{q^F x_{maj}-1}{q^F-1}~,
\end{equation}
where the number of states is now $q^F$, since a state is a vectorial quantity.
In this case, the MF definition and the Monte-Carlo one are slightly different: in fact, here 
we compare the occupations of vectorial states whereas in MF we compare the occupations
 of two traits in a feature. In the following we will denote with $M_{maj}$ the random 
variable whose thermal average per particle gives $m_{MC}$.

The values of $\beta_0$ we use to build the histograms 
are fixed according to Matsuda's analytical predictions \cite{Matsuda81} for $F=2$ 
(see also \App{app:previous}): 
$\beta_c J=0.333136$ for $q=2$, and $\beta_c J=0.401812$ for $q=3$, 
restricting to six decimal digits.

\begin{figure}[t]
\begin{center}
\includegraphics[height=6cm,angle=0]{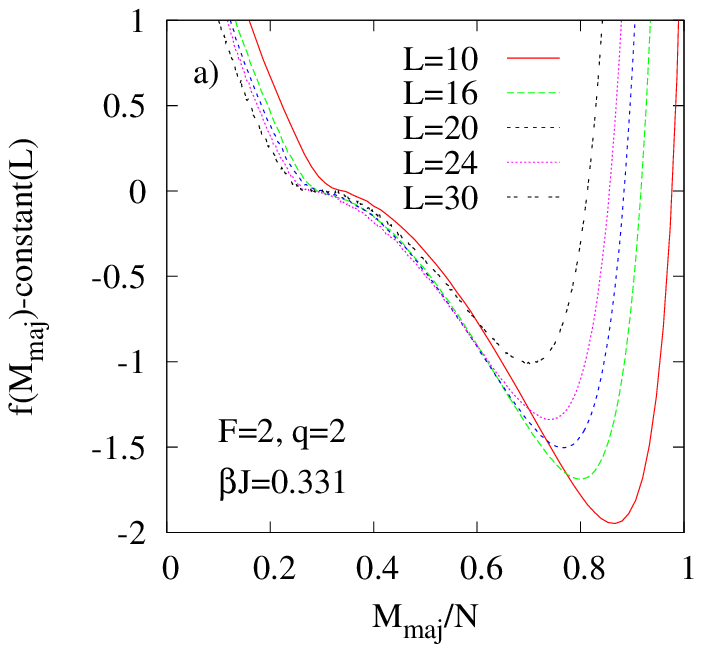}
\includegraphics[height=6cm,angle=0]{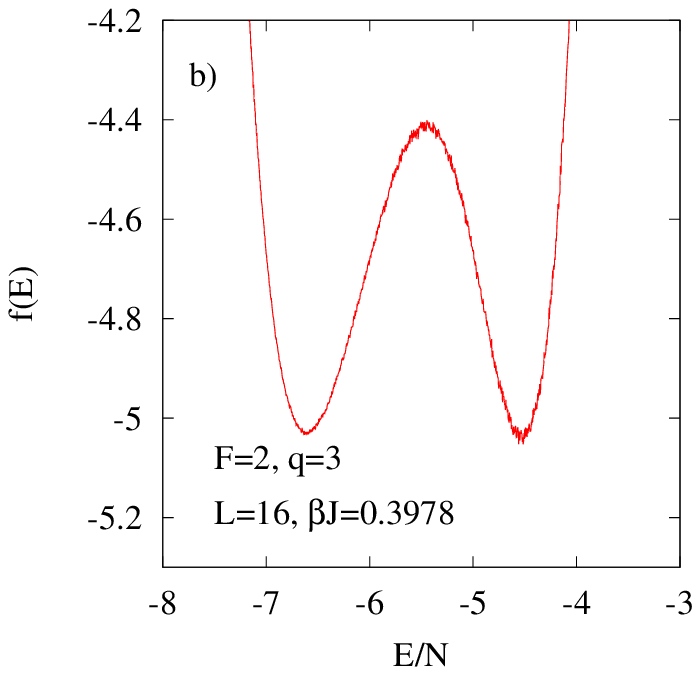}
\end{center}
\caption{Example of the curves from which $\Delta f$ is extracted. 
(a) $f(M_{maj},\beta;L)$ for $F=2,q=2$, $\beta J=0.331$ and five values 
of $L$. The curves have been shifted vertically so that the inflection point
 is on the horizontal axis.
(b) $f(E,\beta;L)$ for $F=2,q=3$, $\beta J=0.3978$ and $L=16$. 
In all the cases, $L\times L$ lattices are used, and $N=L^2$ is the total number of sites.
}
\label{fig:exampleF}
\end{figure}

In \Fig{fig:exampleF},
we show typical shapes of $f$: in \Fig{fig:exampleF}(a), we give $f(M_{maj},\beta;L)$, for the case $F=2,q=2$, for 
five values of $L$ and a given temperature; in \Fig{fig:exampleF}(b), we show $f(E,\beta;L)$, for the case $F=2,q=3$, 
for a given $L$ and temperature.

Let us focus on \Fig{fig:exampleF}(a), $F=2,q=2$. Since at low temperature and in the absence of an external 
field there are $q^F$ degenerate states, in this case we expect four degenerate states.
In this plot, obtained using the majority magnetization, the information about the direction of 
alignment is lost: the global minimum we see corresponds to ordering, in any direction.
To distinguish between orientations in different directions, we could use a bi-dimensional order parameter:
this variable would show indeed four degenerate minima, but the calculation of the 
free energy barrier between them would be very complicated.
Instead, we identify the barrier in this case as the difference between the global minimum and 
the inflection point. At the same time, we take care that the run parameters and its length are such that 
all the degenerate minima are approximately equally visited during the simulation.
As a test, we applied this recipe to case of a 3-state Potts model, and 
we could reproduce, a part from an irrelevant constant, the results of \Ref{Kosterlitz93}.

\begin{figure}[t]
\begin{center}
\includegraphics[height=5.5cm,angle=0]{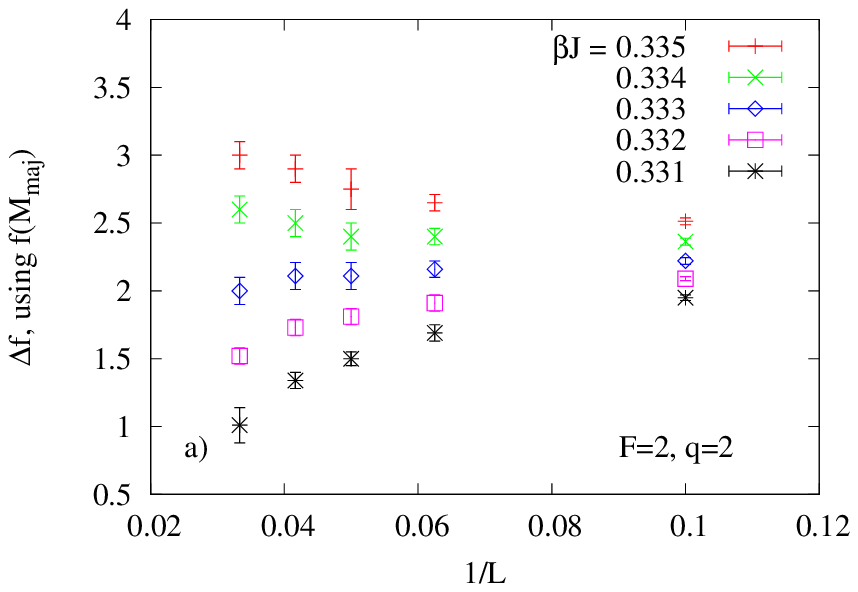}
\includegraphics[height=5.5cm,angle=0]{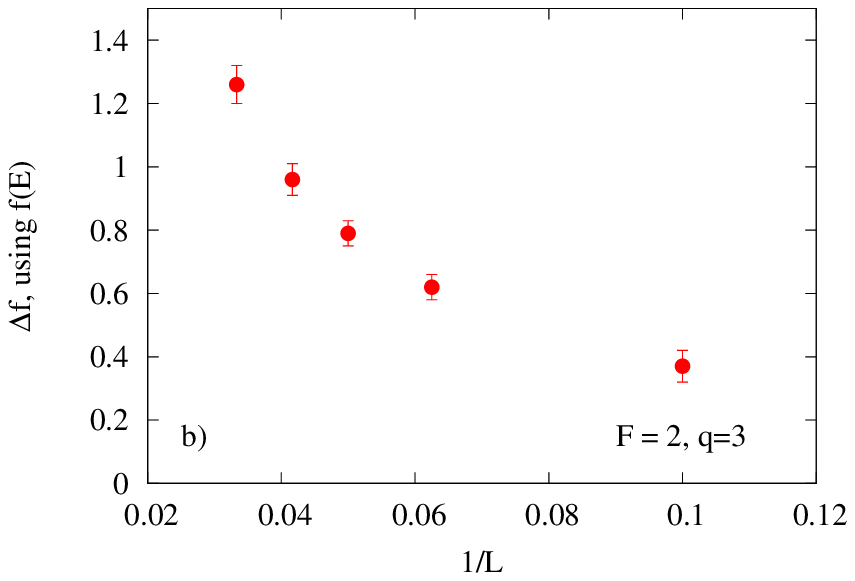}
\end{center}
\caption{Free energy barrier $\Delta f$ as a function of the inverse of the lattice size $L$.
(a) Case $F=2,q=2$, analysis of field-driven phase transitions. 
The barrier $\Delta f$ is obtained from $f(M_{maj},\beta;L)$, for a set of temperature values.
The symbol shape denotes a given value of the temperature.
(b) Case $F=2,q=3$, analysis of temperature-driven phase transitions. 
 The barrier $\Delta f$ is obtained from $f(E,\beta;L)$. For each $L$
 the barrier is computed at the temperature at which the two minima become (roughly) degenerate.
In both panels, $L\times L$ lattices are used, with $L=10,16,20,24,30$.
}
\label{fig:LKF2q23}
\end{figure}

From these and similar figures we extract $\Delta f$ in a range of temperature and sizes: the 
results for $F=2$ are summarized in
\Fig{fig:LKF2q23},
it only remains to interpret them as explained above.
For $F=2,q=2$, from \Fig{fig:LKF2q23}(a) we see that a field-driven phase transition is abrupt for $\beta J>0.333$ 
and seems continuous for $\beta J\simeq 0.333$. Therefore, the temperature-driven phase transition at $\beta J \simeq 0.333$
 appears continuous. For $F=2,q=3$, from \Fig{fig:LKF2q23}(b) we deduce that the temperature-driven phase transition taking place 
at $\beta J\simeq 0.4$ is abrupt.
Since we are interested in temperature-driven transitions, the reader 
may wonder why we do not use the approach of \Fig{fig:LKF2q23}(b) also for $F=2,q=2$. The reason is 
simple: for $F=2,q=2$ we find a single minimum in the restricted free energy $f(E)$. 
While this is already an indication that no jump is present, so, if a transition occurs, 
it is probably continuous, we prefer to resort to another a point of view, that of $f(M)$, and apply the 
Lee-Kosterlitz
method on it. As already discussed above, it is hard to distinguish, numerically, between a weakly 
first-order phase transition and a continuous one. What we can safely state is that 
 the results of \Fig{fig:LKF2q23}(a) are consistent with the presence of a continuous 
phase transition, and doing tests with $L$ up to $120$ we never saw the emergence of a peak in $f(E)$.

We repeat a similar analysis for the case of three coupled layers ($F=3$), and $q=2,3$.
Unfortunately, for $F=3$ we do not have an analytical prediction for the transition temperature for 
the model, \Eq{hamil}. In fact, analytical calculations based on duality arguments 
require the inclusion of six-point terms in the Hamiltonian in the case of three layers (see \Ref{Matsuda83}). 
We then need to estimate the transition temperature: using a standard finite-size scaling 
study \cite{Gould96,Binder00}, we obtain $k_B T_c/J=4.9096(7)$ for $F=3,q=2$; 
for $F=3,q=3$, using the results on a $40\times40$ lattice, we estimate $k_B T_c/J \simeq 4.06$.
The results for $F=3$ are summarized in 
\Fig{fig:LKF3q23}:
\begin{figure}[t]
\begin{center}
\includegraphics[height=5.5cm,angle=0]{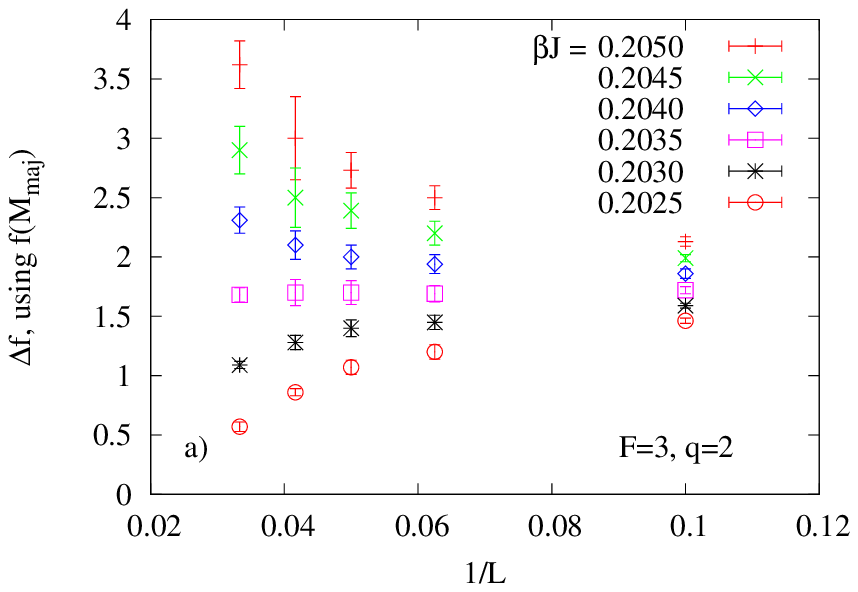}
\includegraphics[height=5.5cm,angle=0]{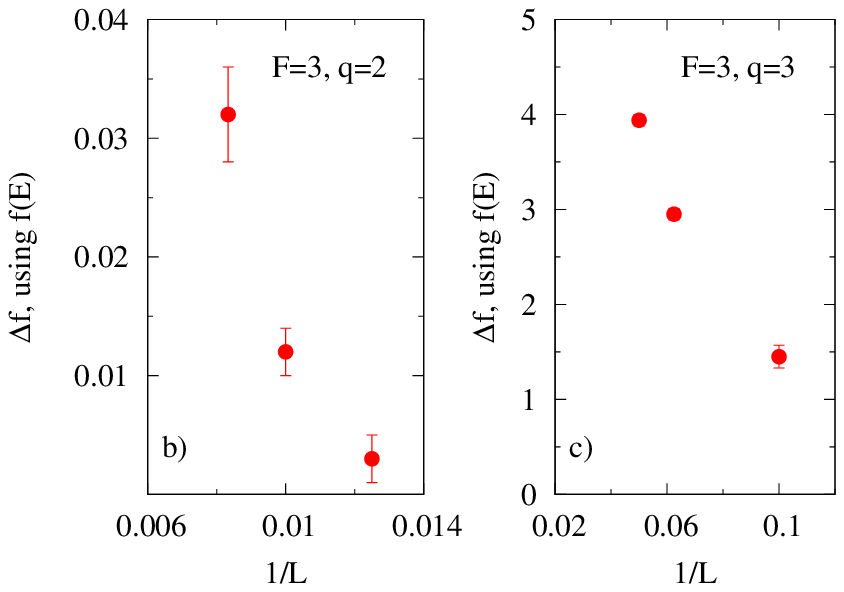}
\end{center}
\caption{Free energy barrier $\Delta f$ as a function of the inverse of the lattice size $L$.
(a) Case $F=3,q=2$, analysis of field-driven phase transitions. 
The barrier $\Delta f$ is obtained from $f(M_{maj},\beta;L)$, for a set of temperature values.
The symbol shape denotes a given value of the temperature. 
Analysis of temperature-driven phase transitions for 
(b) $F=3,q=2$, and (c) $F=3,q=3$.
 The barrier $\Delta f$ is obtained from $f(E,\beta;L)$, as explained in \Fig{fig:LKF2q23}.
 In all the panels, $L\times L$ lattices are used: in (a), $L=10,16,20,24,30$; 
in (b), $L=80,100,120$, and in (c), $L=10,16,20$.
}
\label{fig:LKF3q23}
\end{figure}
the case with $q=2$ in \Figs{fig:LKF3q23}(a) and \ref{fig:LKF3q23}(b)
and the case with $q=3$ in \Fig{fig:LKF3q23}(c).
Again, we find that for $q=3$ an abrupt temperature-driven phase transition takes place 
[see \Fig{fig:LKF3q23}(c)]. For 
$q=2$, and lattices of linear size $L$ up to $30$, we observe the same behaviour found 
for $F=2, q=2$ [compare \Fig{fig:LKF2q23}(a) and \Fig{fig:LKF3q23}(a)].
 However, increasing the lattice size, we find that a two-minima structure 
develops in $f(E,\beta;L)$ for $L\ge 80$: in \Fig{fig:LKF3q23}(b) we show the 
values of the energy barrier as a function of $1/L$. A comparison between \Figs{fig:LKF3q23}(b) 
and \ref{fig:LKF3q23}(c)
shows clearly that the phase transitions for $F=3,q=2$ and $F=3,q=3$, even if qualitatively 
similar, are quantitatively distinct: for $F=3,q=2$, where the phase transition is 
weakly abrupt, it is necessary to increase the 
lattice size by one order of magnitude to be able to see energy barriers which are two orders 
of magnitude smaller than in the case $F=3,q=3$.

Coming to the discussion of the results presented in this section,
first of all, our findings for the phase transition type should be contrasted with the 
known result \cite{Wu82} that the $q$-state Potts model 
in a two-dimensional square lattice undergoes a continuous temperature-driven phase transition 
for $q\le 4$ and an abrupt one for $q>4$. We find that the coupling of a layer with other ones 
(at least for the coupling considered here) results in a lowering of
the value of $q$ for which the transition type changes from continuous to abrupt.
Also, a comparison with the MF results shows that this approximation fails to catch 
the correct transition type only for $F=2,q=2$. 
Concerning the transition temperatures, they are, as usual, overestimated at the MF level,
but the relative error gets smaller and smaller as the total number of states ($q^F$) increases. 
We mention that for the $q$-state Potts model it was conjectured that the MF approximation 
describes accurately the phase transition, in two or more spatial dimensions, when $q$ is large \cite{Mittag74}.

Second, we make a comparison with previous results available in the literature for coupled Potts models 
or similar ones. A summary of the expected nature of the temperature-driven phase transition in $F$ coupled
$q$-state Potts models in a two dimensional square lattice can be found in \Ref{Pujol1996}, 
and references therein. For an attractive coupling between the layers, as in our case, the 
transition type has been studied for different choices of $F$ and $q$: for $F=2$ and $q=2$, 
it is continuous \cite{Baxter} (notice that, in this case, our model coincides with the Ashkin-Teller 
isotropic model, with a relative strength of the four-body term fixed to $K_4/K=1/2$, in Baxter notation);
for $F=2$ and $q>2$, there are indications that it is abrupt \cite{Pujol1996};
for $F>2$, and $q=2$, there are indications that it is abrupt (from \Ref{Grest81}, which studies the $F$-colour AT model
 for weak four-body coupling); in the limit of $F\rightarrow\infty$ and $q=2$, 
it has been shown exactly that it is abrupt (from \Ref{Fradkin1984}, which studies the 
$F$-colour AT model with a four-body coupling $\propto 1/F$).
We recall here, as already discussed in \Sec{Sec:MF}, that in the AT model new phases appear 
when the four-body interaction is larger than the one considered in this work \cite{Baxter,Grest81}. 
The existence of a single phase-transition is also a necessary 
ingredient for Matsuda's predictions based on self-duality: the agreement between our numerical results for the 
transition temperature and his prediction for $F=2$ confirms this point.
More recently, in \Ref{Genzor14}, Genzor {\it et al.} have shown results, obtained using the Corner Transfer Matrix 
Renormalization Group method, for our model, for $F=2$, $q=2,3$ (see Fig. 2 of 
that reference).
While for $q=2$ their results for the transition temperature ($k_B T_c/J=3.0012$) and phase transition 
type are in agreement with ours and Matsuda's ones, for $q=3$ they differ: in fact, in this case 
they find $k_B T_c/J=2.5676$ (Matsuda prediction, shown up to four decimal digits, is $2.4887$) and a continuous 
temperature-driven phase transition.

Finally, keeping in mind the differences between equilibrium and non-equilibrium
models, we would like to recall at this point that for the non-equilibrium
Axelrod model, in the absence of noise, an order-disorder phase transition is
also found. In this case, it takes place as the number of traits $q$ is
varied, at a value $q_c(F)$, and the type of the phase transition changes with
$F$: it is continuous for $F=2$ and abrupt for larger $F$ \cite{Castellano00}. 
When noise (the temperature counterpart in this kind of models)
 is added to the original Axelrod
model, continuous order-disorder transitions controlled by the noise rate
are found in finite systems \cite{Klemm03}. A pseudo-critical noise rate $r_c$
at which the system behaviour changes can be identified \cite{Toral07}: 
for $r>r_c$ the system is disordered,
while for $r<r_c$ it tends to a mono-cultural state.
The quantity $r_c$ depends on the system size $N$, in a way that
depends on the lattice properties; for a regular
two-dimensional network, it is $r_c \approx 1/N \ln N$.
Therefore, in the limit of infinite size $(N\rightarrow \infty$) 
the system is disordered as soon as $r\ne 0$.
On the other side, in our thermodynamical version, that refers to infinite systems,
we always find phase transitions.
Therefore, in contrast with the one-dimensional case \cite{Gandica13},
in two spatial dimensions the thermodynamical version of the Axelrod model
and the out-of-equilibrium original model with noise show
a different qualitative behaviour when $N \rightarrow \infty$.

In the following section, we will address the geometrical properties of the system, 
showing that data collapse can be obtained in the continuous phase transition ($F=2, q=2$),
but also in the weakly first-order one ($F=3$, $q=2$).
\section{Percolation critical behavior}\label{sec:perco}
At the critical point, as a consequence of the divergence of the correlation length, 
important simplifications take place in the thermodynamic properties, and lead to a universal 
(in the sense of the renormalization group) thermodynamic behaviour \cite{kardar,goldenfeld,stanley}.
Moreover, remarkably, in lattice spin models also some geometrical properties are characterized by 
interesting simplifications \cite{stauffer,christensen}.
The divergence of the correlation length causes a power-law dominance, in terms of the relevant scale, in the cluster's
statistics functional form, in the vicinity of the critical point \cite{stauffer,christensen,fortunato02}.

In this Section we study the finite-size scaling of the cluster distribution in our model, to give a geometrical point of view 
on the phase transitions analyzed in the previous section. 
Since for the geometrical properties one faces the same difficulties discussed above in discriminating 
between a continuous phase transition and a weakly first-order one \cite{Peczak89}, we consider here 
both the $F=2,q=2$ and $F=3,q=2$ case. For the latter, in which an abrupt phase transition takes place,
it is intended that the behaviour is only pseudo-critical, and the related exponents are pseudocritical as well.
We focus on geometrical clusters, that is, sets of nearest neighbour sites in the same state, and 
indicate with $s$ the number of sites belonging to the cluster, also called the cluster {\it mass}.
The cluster number density $n(s,T;L)$ gives the average number of clusters of size $s$ that is found at a temperature $T$ 
in a lattice $L\times L$, divided by the total number of sites. 
The finite-size scaling ansatz for the cluster number density at the critical point, $n(s,T_c;L)$, 
states that \cite{christensen}
\begin{equation}\label{eq:scaclus}
n(s,T_c;L) \propto s^{- \tau} \phi(s/L^D ),\quad L\gg 1, s\gg 1~,
\end{equation} 
where the function $\phi$ contains the information about the decay of the cluster number density for sizes 
bigger than the correlation length. 
The constants $\tau$ and $D$ are the Fisher critical exponent 
and the fractal dimension of the incipient infinite cluster (i.e., the largest cluster at $T=T_c$, also called 
the percolating cluster at $T=T_c$), 
respectively, and are related by the hyperscaling law
\begin{equation}\label{eq:hypsca}
\tau=\frac{d}{D}+1~, 
\end{equation}
where $d$ is the spatial dimensionality.
\Eq{eq:scaclus} implies that we can collapse the curves for the cluster 
number density corresponding to lattices of different sizes, if we plot the transformed cluster number density, 
$s^\tau n(s,T_c;L)$, against the rescaled cluster size, $s/L^D$ \cite{christensen}.

Since $\tau$ can be obtained from $D$ using \Eq{eq:hypsca}, 
to perform the collapse we only need to determine the fractal dimension, $D$, of the percolating cluster at the critical temperature.
For this, we measure how the percolating cluster fills the space when the correlation length diverges, in terms of the
lattice linear size, $S_{\infty} \varpropto L^D$ \cite{christensen}. In our calculations, we measure the mass of the percolating 
cluster by means of the size of the biggest cluster at $T_c$.
One early signal related to the kind of phase transition is the error in the
linear fit for the log-log plot in the determination of $D$. The cases $F=2,3$ and $q>2$ as well as $q>5$ in the Potts model
give high values for the Asymptotic Standard Error. On the contrary, we get a very low Standard Error $(< 0.2 \%)$ when $q=2$
for $F=2$ and $3$. Again, for the considered lattice sizes, the phenomenology of $F=2,q=2$ and $F=3,q=2$ is 
similar. Using lattices of size up to $L=100$, we obtain for the fractal dimension the values $D=1.885(2)$ for $F=2$ and 
$D= 1.882(2)$ for $F=3$. The corresponding values of the Fisher exponent are $\tau=2.061(2)$ 
and $\tau=2.063(2)$ for $F=2$ and $3$, respectively.

\begin{figure*}[t]
\begin{center}
\scalebox{0.7}{\includegraphics{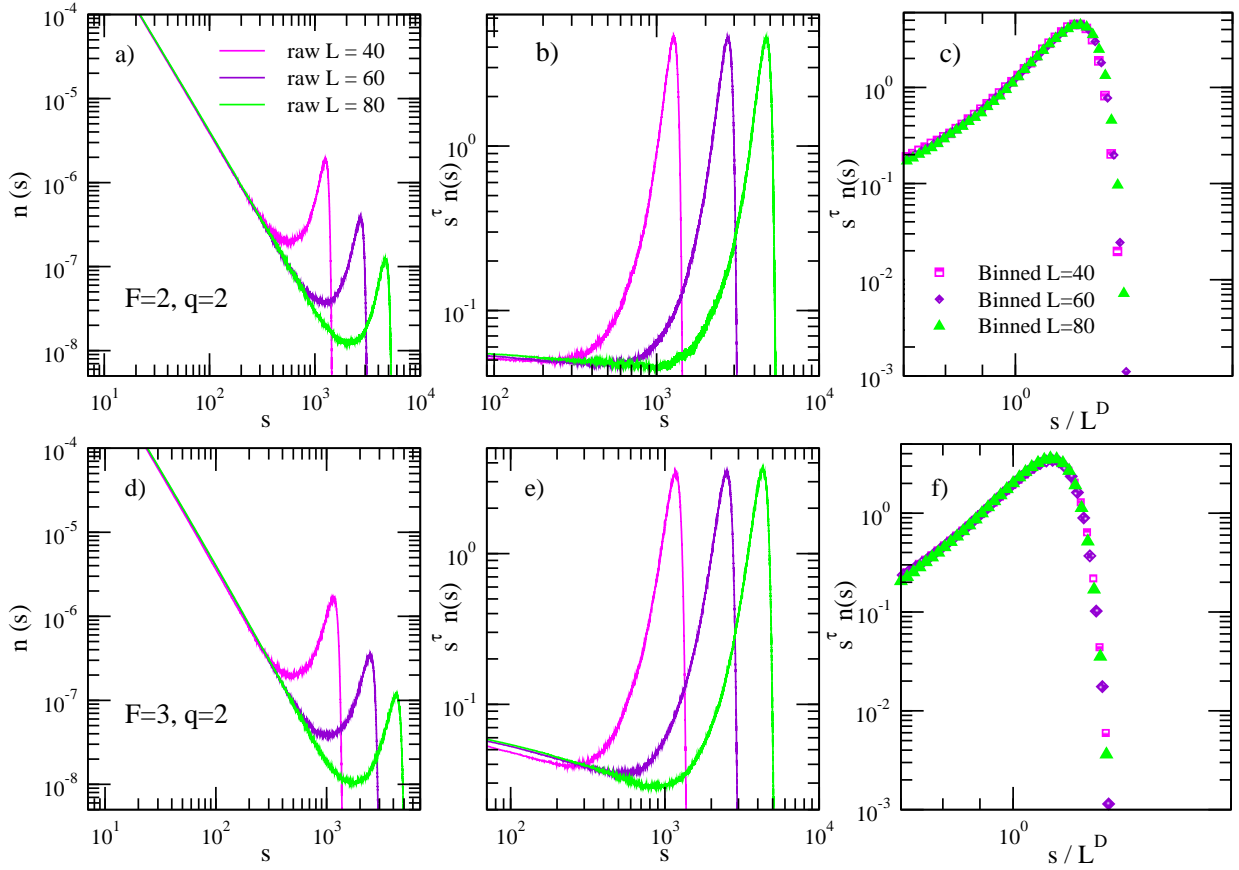}}
\end{center}
\caption{Data collapse for the cluster number density, at $T=T_c$, for sizes $L= 40$, $60$ and $80$. 
(a)-(c) The critical case $F=2, q=2$ and (d)-(f) the pseudo-critical case $F=3, q=2$. 
We show (a) and (d) the raw data for the cluster number density $n(s)$ against $s$, 
(b) and (e) the raw data for the transformed cluster number density $s^\tau n(s,T_c;L)$ 
against $s$, and (c) and (f) the binned version of the transformed cluster number density 
$s^\tau n(s,T_c;L)$ against the rescaled argument $s/L^D$. }
\label{collapse}
\end{figure*}
In \Fig{collapse} we show the construction of the data collapse for the cluster number density as a function of
the cluster size, for two values of $F$ and $q=2$. The critical case $F=2$ is shown in the top panels
 and the pseudo-critical case $F=3$ is given in the bottom panels. 
Figures \ref{collapse}(a) and \ref{collapse}(d) show the raw data for the cluster number density, i.e, the normalized cluster size distribution against 
$s$, for $L=40, 60$ and $80$. The scale invariance over more
than $5$ order of magnitude is evident and one can notice, at the same time, the appearance of the incipient infinite clusters 
(at larger sizes). Figures \ref{collapse}(b) and \ref{collapse}(e) display the transformed cluster number density $s^\tau n(s,T_c;L)$ against the cluster size,
 $s$. Then, the onset of the rapid decay is in the same vertical position. To achieve the data collapse, 
the clusters size should be rescaled by the characteristic cluster size,
$s_\xi$, which goes as $\varpropto \xi^D$ in infinite systems \cite{christensen}. However, when systems are finite, the lattice sizes
are always less than the correlation length and hence limit the characteristic cluster size: as a consequence
$s_\xi \varpropto L^D$. Figures \ref{collapse}(c) and \ref{collapse}(f) depict the binned data for the collapse of the curves once the
rescaling of the horizontal axis 
$s \rightarrow s/L^D$ has also been performed, 
illustrating the perfect collapse. 
Moreover, our data collapse also confirms the value of the critical exponent $\tau$, and of $D$.

Finally, we recall that for the case $F=1,q=2$ (2-state Potts model), the value of $D$ predicted by Vanderzande 
\cite{Vanderzande92} is $187/96=1.9479\dots$:
we find therefore that when coupling two or three layers the fractal dimension $D$ is altered. The results 
for $F=2$ and $F=3$, however, are compatible within their error bars.
As a test, we also compute $D$ for the case $F=1$, and in our simulations we get $D=1.9458(8)$, 
which agrees with the prediction by Vanderzande up to the second decimal. This indicates that  
the errors from the fits, which we have given in parenthesis, underestimate the real error, which 
is most likely in the second decimal digit. However, even with a conservative estimate 
of the error, our findings on the variation of $D$ are confirmed: $D=1.89$ for $F=2$, while $D=1.95$
 for $F=1$. For $L$ up to $100$, $F=3$ has a pseudo-critical behaviour 
with $D=1.88$.
\section{Conclusions}\label{sec:conclu}
We have presented in this work a study of $F$ coupled $q$-state Potts models in a two dimensional
 square lattice, with an attractive four-body inter-layer coupling of fixed strength.
 We have analyzed the nature of phase transitions as the number 
of internal degrees of freedom $q$ and the number of layers $F$ are varied. 
Using the Lee-Kosterlitz method, we have found that 
the temperature-driven phase transition in zero field is continuous for $F=2,q=2$ 
and abrupt for higher values of $q$ and/or $F$.
At mean-field level, we have found that for $F\ge 2, q\ge2$ all the zero-field phase transitions are abrupt. 
This effect is a typical consequence of the suppression of fluctuations,
 whose strength is responsible for the singularities that characterize second-order 
phase transitions. However, the mean-field approximation has indicated the general dependence 
of the transition temperature upon $F$ and $q$, confirmed by MC results. The accuracy of the MF 
predictions for these temperatures improves as the total number of states ($q^F$) increases.

For $F=2,q=2$ and $F=3,q=2$, we have studied the properties of the geometrical clusters at the critical point, 
{\it via} finite-size scaling and data collapse. This has allowed us to determine the fractal dimension
 of the percolating cluster, showing that, when two layers are coupled,
this fractal dimension is reduced from $D\simeq 1.95$ to $D\simeq 1.89$.
When three layers are coupled, a pseudo-critical behaviour is found, with
$D\simeq 1.88$.

We have made comparisons with previous results from the field of statistical mechanics, 
and, also, with the results obtained for the Axelrod model of social dynamics, a non-equilibrium 
model with some similar characteristics. In contrast with the one-dimensional case,
in two spatial dimensions we have found that the thermodynamical version of the Axelrod model
and the out-of-equilibrium original model with noise show 
a different qualitative behaviour in the limit of infinite size.
This comparison is an example of the possible interesting connections that can be drawn between 
statistical mechanics and complex systems.
\appendix
\section{Mean-Field approximation for $F=1$ --- The q-state Potts model}\label{app:MFpotts} 
We present here, closely following \Ref{LoicTurban}, the MF approximation for the simple case of one layer,
 on which the case of generic $F$ is based. The calculations are done for a generic lattice, with coordination number $z$. 
The Hamiltonian (\ref{hamil}) reduces in this case to the $q$-state Potts model
\begin{equation}
\mathcal{H}_\potts=-J\sum_{\nen}\delta(n_i,n_j)-H\sum_i\delta(n_i,0)~,
\end{equation}
where $n_i=0,1,\dots,q-1$.
Without loss of generality, we have assumed that the state $0$ is the favored one.
We need to define two quantities: the probability $p({n_i})$ that a given site 
is in the state $n_i$, and the order parameter variable $s({n_i})$, whose average 
will describe the phase transition. The averages at one site $i$ are computed as 
$\langle A\rangle_i=\sum_{n_i=0}^{q-1}p(n_i)A(n_i)$. 

In a MF approximation, the whole probability factorizes as 
\begin{equation}
p_{MF}(n_1,n_2,\dots,n_N)=\prod_{i=1}^N p_i(n_i)~.
\end{equation}
We make the following Ansatz: the single-site probability is
$p(n_i)=a + b\,\delta(n_i,0)$ and  $s(n_i)=c + d\,\delta(n_i,0)$, where $a,b,c,d$ are 
constants. The parameters $a$ and $b$ are fixed normalizing the 
probabilities ($\sum_{n_i=0}^{q-1}p(n_i)=1$) and imposing that the average of the order 
parameter is $m$. The parameters $c,d$ are found imposing that $\langle s\rangle_i=1$ for $T=0$,
and $\langle s\rangle_i=0$ for high temperatures. 

One finds:
\begin{equation}
p(n_i)=\frac{1+m[q\, \delta(n_i,0)-1]}{q}~,\quad n_i=0,1,\dots,q-1.
\end{equation}
and 
\begin{equation}
s(n_i)=\frac{q\, \delta(n_i,0)-1}{q-1}~,\quad n_i=0,1,\dots,q-1.
\end{equation}

To better understand the meaning of the quantity $m$, notice that it is the difference between 
the average occupation of state $0$ and of a state different from state $0$.
In fact: the average occupation of state $0$ is $\mu\equiv\langle \delta(n_i,0)\rangle_i=\frac{1+m(q-1)}{q}=m+\frac{1-m}{q}$
 and for state $1$ (and any state other than $0$) it is $\nu\equiv\langle \delta(n_i,1)\rangle_i=\frac{1-m}{q}$. Clearly, $\mu-\nu=m$.

Using the following exact relation
\begin{equation}\label{eq:exdeltas}
\delta(n_i,n_j)=\sum_{\alpha=0}^{q-1}\delta(n_i,\alpha)\delta(n_j,\alpha)~,
\end{equation}
we rewrite the Hamiltonian in a form which allows to evaluate directly the averages over single sites
\begin{eqnarray*}
&&\langle \mathcal{H}_\potts \rangle_{MF}=\nonumber\\
&=&-J \sum_{\nen}\sum_{\alpha=0}^{q-1}\langle\delta(n_i,\alpha)\rangle_i
\langle\delta(n_j,\alpha)\rangle_j-H\sum_i\langle\delta(n_i,0)\rangle_i\\
&=&-J\frac{Nz}{2}[\mu^2+(q-1)\nu^2]-NH\mu\\
&=&-J\frac{Nz}{2q}[1+(q-1)m^2]-\frac{NH}{q}[1+m(q-1)]~.
\end{eqnarray*}

In the above equality we have used the relation $\mu^2+(q-1)\nu^2=[1+m^2(q-1)]/q$ that one obtains 
rewriting $\mu$ and $\nu$ in terms of $m$.

For the statistical entropy term we have:
\begin{eqnarray*}
\langle \ln{p_{MF}}\rangle_{MF} &=& \sum_{i=1}^N \langle \ln{p_i} \rangle_i 
= \sum_{i=1}^N \sum_{n_i=0}^{q-1}p(n_i) \ln{p(n_i)}\\
&=& N\left[\mu \ln{\mu} +(q-1) \nu \ln{\nu}\right]\\
&=& N\bigg\{\frac{1+m(q-1)}{q}\ln \frac{1+m(q-1)}{q}\\
&&+(q-1)\frac{1-m}{q}\ln{\frac{1-m}{q}}\bigg\}~.
\end{eqnarray*}

Finally, the Gibbs free energy per particle of the $q$-state Potts model in the MF approximation is 
\begin{eqnarray}\label{eq:MFpotts}
g_{MF}^\potts(m,T)&\equiv&\frac{G_{MF}}{N}=\frac{\langle \mathcal{H}_\potts \rangle_{MF} +k_BT\langle \ln{p_{MF}}\rangle_{MF}}{N} \nonumber\\
&=&-J\frac{z}{2q}[1+(q-1)m^2]-\frac{H}{q}[1+m(q-1)]\nonumber\\
&&+k_BT 
\bigg\{\frac{1+m(q-1)}{q}\ln \frac{1+m(q-1)}{q}\nonumber\\
&&+(q-1)\frac{1-m}{q}\ln{\frac{1-m}{q}}\bigg\}~.\nonumber\\
\end{eqnarray}
\subsection*{Analytical results for the MF approximation}
For this $F=1$ case, analytical results for the transition temperature and type
can be obtained at the MF level \cite{Wu82}.
The inverse transition temperature $K_c=J/k_BT_c$ is
\begin{equation}
K_cz =2\frac{q-1}{q-2}\ln{(q-1)}
\end{equation}
and the jump in the order parameter at transition is 
\begin{equation}
m_c=\frac{q-2}{q-1}~.
\end{equation}
The latter equation shows that at the MF level the phase transition is continuous 
for $q=2$ and abrupt for $q>2$.
\section{Previous exact results for the two-dimensional square lattice}\label{app:previous}
Using self-duality arguments, the transition temperature of our model 
on a square lattice has been previously located 
for $F=1$ (Potts model, see, e.g. \cite{Wu82}) and for $F=2$ \cite{Matsuda81}.

For $F=1$ and generic $q$, our model is a standard $q$-state Potts model whose transition in a square lattice
is located solving
\begin{equation}
e^{K_c}=1+\sqrt{q}\Leftrightarrow K_c=\ln(\sqrt{q}+1)~,\quad (F=1).
\end{equation}

For $F=2$, generic $q$, and no external field, our model is equivalent to Matsuda's model \cite{Matsuda81}
with an appropriate choice of couplings. In more detail, our $J$ is related to $J_i$, $i=1,2,3$ of \Ref{Matsuda81} 
as follows: $J_1=J_2=J$ and $J_3=2J$. 
The transition temperature is found \cite{Matsuda81} by solving
\begin{equation}
e^{4K_c}-2e^{K_c}-(q-1)=0~,\quad (F=2).
\end{equation}

\begin{acknowledgments}
The authors are very grateful to Prof. Fernando Sampaio Dos Aidos, Prof. Ernesto Medina 
and Dr. Isaac Vida\~{n}a for valuable discussions. 
They are also indebted with the anonymous referee for helping them to improve the manuscript.
The work of Y.G. presents research results of the Belgian Network DYSCO (Dynamical Systems, Control, and Optimization), funded by the Interuniversity 
Attraction Poles Programme, initiated by the Belgian State, Science Policy Office. 
S. C. is supported by the ``Funda\c{c}\~{a}o para a Ci\^{e}ncia e a Tecnologia'' (FCT, Portugal) and the 
``European Social Fund'' (ESF) {\it via} the postdoctoral grant No. SFRH/BPD/64405/2009.
\end{acknowledgments}

\end{document}